\documentclass[aps,prl,reprint,superscriptaddress]{revtex4-2}

\usepackage{graphicx}
\usepackage{dcolumn}
\usepackage{bm}
\usepackage{natbib}
\usepackage{hyperref}
\usepackage{float}
\usepackage{xcolor}
\usepackage[normalem]{ulem}

\begin{document}


\title{Tailoring 4H-SiC Surface Electronic States by Atomic-Layer Deposition for Ideal Peta-Ohm Resistors}


\author{Yuying Xi}
\thanks{The two authors contribute equally to this work}
\affiliation{College of Electronic Information and Optical Engineering, Key Lab of Interface Science and Engineering in Advanced Materials, Key Lab of Advanced Transducers and Intelligent Control System of Ministry of Education, Taiyuan University of Technology; Taiyuan, 030024, China}

\author{Helios Y. Li}
\thanks{The two authors contribute equally to this work}
\affiliation{Department of Mechanical Engineering, The University of Hong Kong; Hong Kong, 999077, China}

\author{Guohui Li}
\email{liguohui@tyut.edu.cn}
\affiliation{College of Electronic Information and Optical Engineering, Key Lab of Interface Science and Engineering in Advanced Materials, Key Lab of Advanced Transducers and Intelligent Control System of Ministry of Education, Taiyuan University of Technology; Taiyuan, 030024, China}

\author{Qingmei Su}
\affiliation{School of Materials Science and Engineering and Materials Institute of Atomic and Molecular Science, Shaanxi University of Science and Technology; Xi'an, 710021, China}

\author{Kaili Mao}
\affiliation{Shanxi Semisic Crystal Co., Ltd.; Taiyuan, 030000, China}

\author{Bingshe Xu}
\affiliation{College of Electronic Information and Optical Engineering, Key Lab of Interface Science and Engineering in Advanced Materials, Key Lab of Advanced Transducers and Intelligent Control System of Ministry of Education, Taiyuan University of Technology; Taiyuan, 030024, China}
\affiliation{Shanxi-Zheda Institute of Advanced Materials and Chemical Engineering; Taiyuan, 030032, China}

\author{Yuying Hao}
\affiliation{College of Electronic Information and Optical Engineering, Key Lab of Interface Science and Engineering in Advanced Materials, Key Lab of Advanced Transducers and Intelligent Control System of Ministry of Education, Taiyuan University of Technology; Taiyuan, 030024, China}

\author{Nicholas X. Fang}
\email{nicxfang@hku.hk}
\affiliation{Department of Mechanical Engineering, The University of Hong Kong; Hong Kong, 999077, China}

\author{Yanxia Cui}
\email{yanxiacui@tyut.edu.cn}
\affiliation{College of Electronic Information and Optical Engineering, Key Lab of Interface Science and Engineering in Advanced Materials, Key Lab of Advanced Transducers and Intelligent Control System of Ministry of Education, Taiyuan University of Technology; Taiyuan, 030024, China}
\affiliation{Shanxi-Zheda Institute of Advanced Materials and Chemical Engineering; Taiyuan, 030032, China}


\date{\today}

\begin{abstract}
High resolution resistors capable of detecting minuscule currents are vital for advanced sensors, but existing off-shelf models struggle with inconsistent resistance under varying voltages. The underlying physics of this issue is rooted in unstable surface charges and intrinsic inhomogeneity of surface potential caused by spontaneous polarization (SP) in commercial semi-insulating silicon carbide (SiC) devices. In this work, we found that coating SiC surfaces with an ultrathin zinc oxide layer immobilizes the dangling surface charges in place and balances the natural electric field of the material, ensuring stable resistance even at extreme voltages up to 1000 V.  The resulting SiC resistor maintains a record-high resistance of one peta-ohm ($\rm 10^{15} \Omega$ ) with negligible voltage fluctuations, outperforming conventional options. Additionally, these devices can switch states when exposed to light or heat, making them dual-purpose tools for ultra-sensitive measurements and sensors. This breakthrough combines high stability, scalability for mass production, and multifunctionality, opening doors to next-generation precision technologies in fields like quantum sensing and environmental monitoring.
\end{abstract}


\maketitle


The demand for atto-ampere-level current detection in emerging technologies \cite{Hashem2011Picoamp, Guglielmi2020ResistorVoltage}, such as biological single-molecule sensing \cite{Li2019Singlemolecule, Sandhu2007Singlemolecule,Tang2022Plasmonic,2022Self} and radiation monitoring \cite{Li2021RadiationDetect}, has led to the need for ultrahigh-resistance resistors (i.e., $>$ 1 $\rm{T\Omega}$) that can convert minute currents into measurable voltages \cite{Park2020OhmtoVoltage, Wang2021OhmtoVoltage, Scarsella2023OhmtoVoltage}. Silicon carbide (4H-SiC), a third-generation semiconductor, combines semi-insulating properties with scalability via 8-inch wafer production \cite{Yang2024,Yang2023}. However, untreated 4H-SiC resistors exhibit voltage-dependent resistance due to their hexagonal crystal structure. The Si- (0001) and C-terminated (000-1) surfaces host mobile charges from dangling bonds, while asymmetric Si-C bonding induces spontaneous polarization \cite{Itoh1997SP, Ji2024SP, Beattie2021SPCharges, Choi2015SP, Huang2020surfacestateSPFermipinning, Pennington2009Trapstates}. This creates imbalanced surface potential, distorting \textit{I-V} behavior.

In this work, we demonstrate that depositing an atomically thin zinc oxide (ZnO) layer on 4H-SiC eliminates voltage-dependent resistance. Experiments reveal two synergistic mechanisms: (1) ZnO immobilizes surface charges via downward band bending below the Fermi level, and (2) its thickness-tuned spontaneous polarization (SP) counteracts 4H-SiC’s intrinsic SP, neutralizing bias across the device. These effects enable resistors with strictly linear \textit{I-V} curves crossing the origin (Fig. 1(c)) and record-high peta-ohm resistances (up to  2 $\rm {P\Omega}$). By engineering surface electronic states, we realize ideal Ohmic ultrahigh resistance on 4H-SiC resistors that meet the long-standing precision demands of atto-ampere sensing, radiation monitoring, and quantum technologies.


\begin{figure*}
\includegraphics[width=14 cm,page=1]{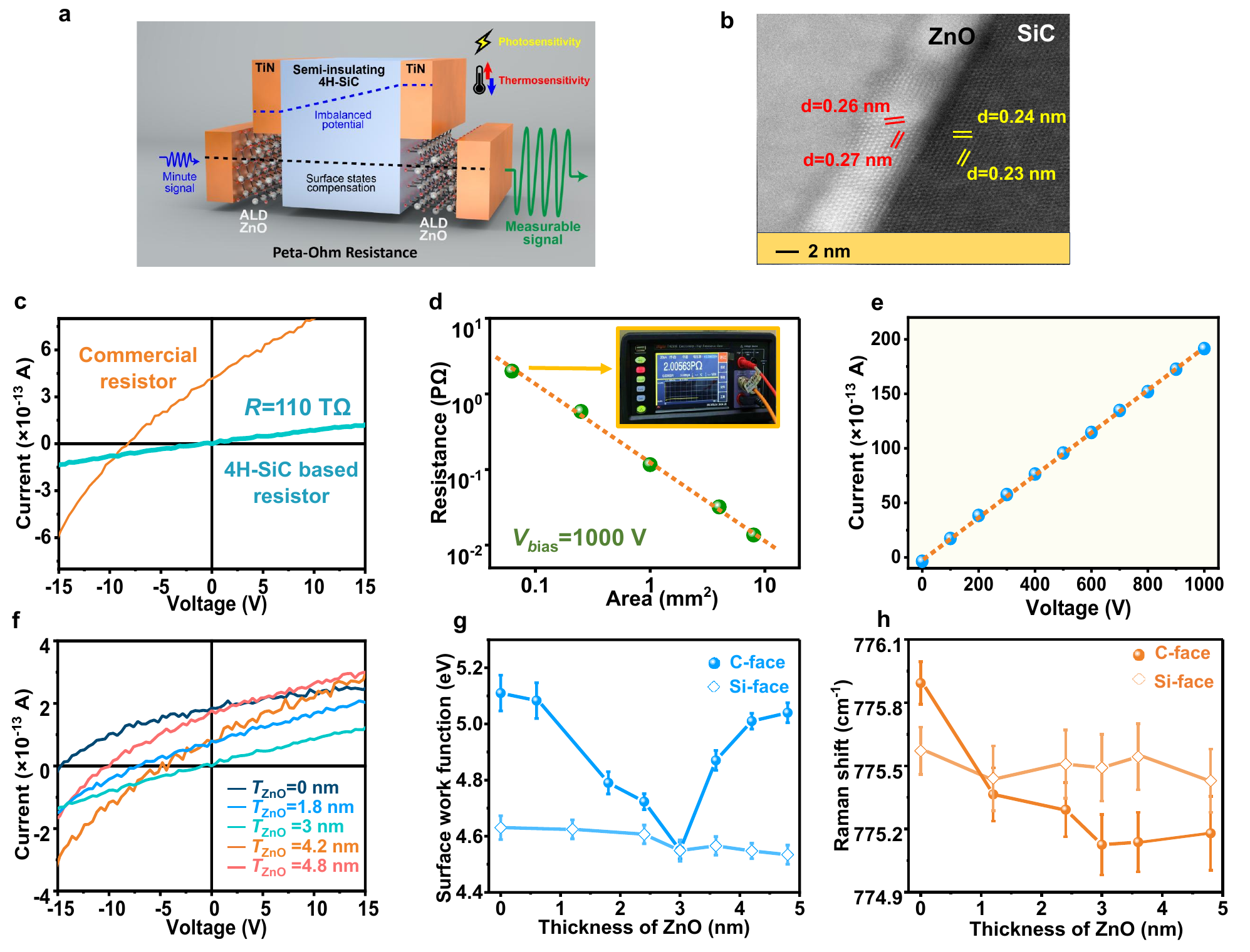}
\caption{\textbf{Properties of 4H-SiC resistors with different thickness of ZnO.} (a) Schematic of the atomically engineered 4H-SiC resistor (ZnO interlayers, TiN electrodes) demonstrating precision measurement via high-resistance amplification of weak current signals into quantifiable voltage signals. (b) In-situ cross-sectional TEM image near the interface of the 4H-SiC/ZnO resistor structure on the C-face. (c) \textit{I-V} measurement comparing 4H-SiC resistor and commercial resistor. (d) Linear dependence of resistance on cross-sectional area for 4H-SiC resistor (inset: resistance testing image of an area of 0.0625 $\rm mm^2$). (e) \textit{I-V} measurement of a 52.4 $\rm T\Omega$ 4H-SiC resistor under bias from 0 to 1000 V. (f) Multiple sets of \textit{I-V} measurement for 4H-SiC resistors with increasing ZnO thickness on the C-face, with a 3 nm ZnO on the Si-face. (g) ZnO thickness dependent surface work functions on the C-face and Si-face. (h) ZnO thickness dependent Raman shift on the C-face and Si-face.}
\end{figure*}

Force distance curves of untreated 4H-SiC measured by atomic force microscopy verified the difference in surface-tip interaction strength (Fig. S1), that is, different surface electronic structures\cite{2023Review}. This supports the facts of imbalanced surface mobile charges induced by dangling bonds in Si- and C- terminated surfaces \cite{Yang2022reasonofsurfacestateSP, Choi2015SP}. Moreover, a slight charge density imbalance in Si-C bonds was produced as a result of the unequal electronegativity between Si and C atoms, introducing a directional intrinsic electric field, that is, SP. This produces an imbalanced surface potential and dipole moment on the Si- and C-terminated surfaces \cite{Itoh1997SP, Ji2024SP}. The combination of uneven surface electronic structures and imbalanced surface potential eventually affects \textit{I-V} curves of 4H-SiC resistors, shown as deep blue nonlinear curves in Fig. 1(f).  This behavior corresponds to a pronounced voltage-dependent resistance characterized by a voltage coefficient of 0.67\%/V, whereby a 10 V voltage variation results in a 6.7\% nonlinear error accumulation, while a 100 V variation escalates the nonlinear error accumulation to 67\%, thus posing a significant challenge to realize precise measurements.

Our strategy of ZnO treatment to 4H-SiC resistors does eliminate the voltage-dependent resistance characteristic. 
The proposed resistor comprises a semi-insulating 4H-SiC wafer, atomic-scale ZnO interlayers through atomic layer deposition (ALD) \cite{Beh2017ALD}, and titanium nitride (TiN) electrodes through magnetron sputtering, as illustrated in Fig. 1(a).
4H-SiC resistors with 3 nm thick ZnO layers on the C-face and Si-face (figure of TEM at C-face in Fig. 1(b)) exhibit a constant resistance across varying voltages, achieving ideal Ohmic behavior and resistance of 110 $\rm T\Omega$ with a cross-sectional area of 1 $\rm mm^2$ and a thickness of ~500 $\rm \mu m$ (Fig. 1(c)). 
The 4H-SiC resistor, with resistance varying from 2 $\rm {P\Omega}$ to 13.6 $\rm T\Omega$ as its cross-sectional area increases from 0.0625 $\rm mm^2$ to 8 $\rm mm^2$(Fig. 1(d)), maintains perfect  \textit{I-V} linearity (Fig. S2) and offers better volumetric integration than commercial resistors (Fig. S3).  
As a wide-bandgap semiconductor with a breakdown strength of $\rm \sim MV/cm$, 4H-SiC enables the resistor to maintain perfect linear Ohmic performance under 1000 V bias, as shown in Fig. 1(e). 
In addition, the 4H-SiC resistors maintain ideal zero-crossing \textit{I-V} curve up to 60 °C, unlike commercial and sapphire-based ones (Fig. S4 and Fig. S5).  See Materials and Methods for more details in materials preparation,  devices fabrication and characterization \cite{sm}.

The ZnO interlayers significantly impact the performance of the resistors. Fig. S6 shows the \textit{I-V} curves of devices with and without ZnO on both the Si and C faces. Devices lacking ZnO modification exhibit nonlinear behavior akin to conventional commercial resistors. Monofacial ZnO coating (3 nm) on the Si-face yields no measurable improvement in performance, still exhibiting nonlinear property. In contrast, unilateral C-face ZnO modification (3 nm) induces zero-crossing Ohmic performance with much suppressed nonlinearity, establishing a direct correlation between interfacial engineering on C-face and \textit{I-V} properties. The inferior linearity performance of single-side C-face modification compared to bilateral configuration suggests unresolved interfacial charge mobilization on the unmodified Si-face. It is found that the voltage offset from 0 V at zero-current conditions systematically changes with the ZnO layer thickness ($T_{ZnO}$) on the C-face (Fig. S7 and Fig. S8). The voltage offset decreases initially and then increases, with the offset reaching zero at a 3 nm thickness of the ZnO layer, further suggesting that this thickness eliminates additional bias caused by surface electronic charges on the C-face. Similarly, ideal \textit{I-V} curves can also be achieved on a thinner 420 $\rm \mu m$ 4H-SiC substrate. As shown in Fig. S9, due to changes in the device's surface properties, the optimal thickness of the ZnO layer for the 420-$\rm \mu m$-thick 4H-SiC was determined to be 1 nm in experiment.

Kelvin probe force microscopy (KPFM) on both C-face and Si-face was performed to evaluate the impact of the ZnO interlayer with varying \textit{T}\textsubscript{ZnO}, with findings illustrated in Fig. S10 and Fig. 1(g). The KPFM measured surface work function exhibits substantial spatial variation across different positions, attributed to uneven doping or surface defects (Fig. S11). To spread out the deviation, we specifically selected a particular sample region to conduct the ZnO-thickness dependent investigation. Depositing a 3 nm ZnO layer reduced the work function from 5.11 eV to 4.55 eV, with further thickness adjustments reversing this trend on C-face, while remaining around 4.6 eV on Si-face (Fig. 1(g)).  Raman spectroscopy detected shifts in the surface electron states on the C-face due to ZnO thickness changes, marked by alterations in the $E_2$ transverse optical phonon peak (from 775.9 $\rm cm^{-1}$ to 775.1 $\rm cm^{-1}$), with no significant change on the Si-face (Fig. 1(h) and Fig. S12). Similarly, Fourier transform infrared spectroscopy identified a C-face-specific interfacial bond shift (931 $\rm cm^{-1}$ to 933 $\rm cm^{-1}$ ) after 3 nm ZnO deposition (Fig. S13), confirming modified chemical interactions. These results selectively demonstrate that ZnO tunes the electronic structure at the C-face, responsible for the ideal Ohmic behavior of the modified 4H-SiC resistor. In addition to ZnO, our investigations extended to alumina ($\rm Al_2O_3$) and titanium oxide ($\rm TiO_2$) interlayers (Fig. S14). Among these, $\rm Al_2O_3$ was identified as an equally effective alternative for modifying the surface states of 4H-SiC, whereas the $\rm TiO_2$ modification did not exhibit any ideal Ohmic performance. See more details of surface work function measurement in Materials and Methods \cite{sm}. 

\begin{figure}
\includegraphics[width=8.6 cm,page=2]{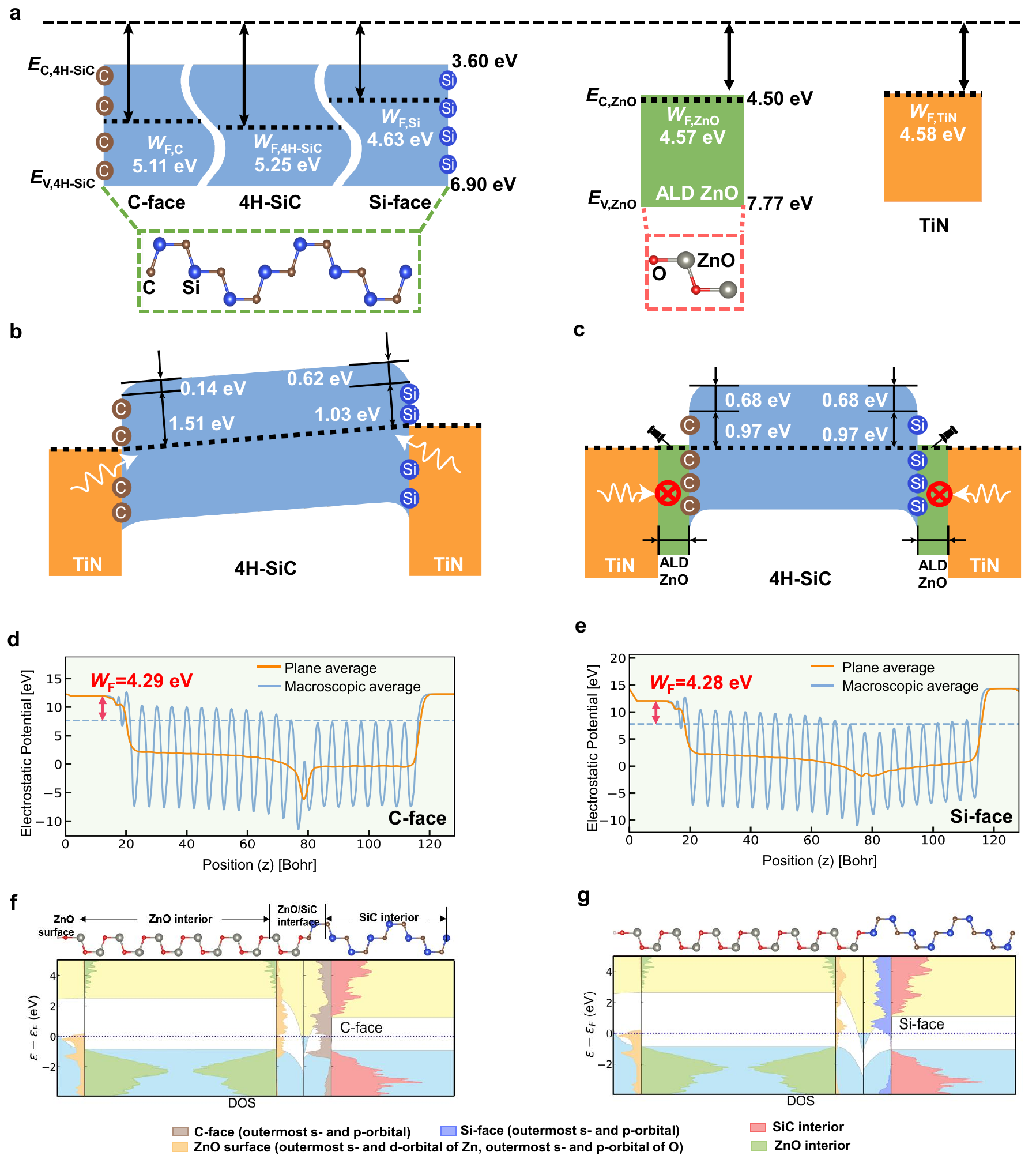}
\caption{\textbf{Interfacial modulation mechanism and DFT simulations of 4H-SiC with both sides.} (a) Energy level diagram of 4H-SiC, ALD ZnO, and TiN. The dashed boxes depict the atomic configurations of 4H-SiC and ZnO. (b) Band diagram of pure 4H-SiC resistors with TiN electrodes demonstrating conductor/semiconductor contact. (c) Band diagram of 4H-SiC resistors with 3 nm ZnO interlayers. (d, e) DFT computed electrostatic potential distribution on the C-face (left) and Si-face (right), dashed lines represent the Fermi level of the structure. $W_{F}$ denotes the potential difference between the vacuum level near the surface and the Fermi level. (f, g) Band diagram of the C-face (left) and Si-face (right) of 4H-SiC with 3 nm ZnO interlayer.}
\end{figure}

Density Functional Theory (DFT) calculations were conducted for gaining physical insights into balancing of surface electronic states of the 4H-SiC resistors, utilizing the atomic configurations of 4H-SiC and ZnO as depicted in Fig. 2(a) for modeling. Both thick slab model simulations (Fig. S15(a)) and DFT simulations of individual C-face and Si-face (Fig. S16) show that the work function of the C-face is consistently higher than that of the Si-face, with both sets of results aligning with the experimental measurement (Fig. 1(g)).  The discrepancies between theoretical and experimental results were attributed to the complex surface states of 4H-SiC during processing; see Materials and Methods for more details \cite{sm}. In principle, the existing SP causes imbalanced surface potential at the terminated two surfaces of a 4H-SiC slab, thus resulting in different work functions \cite{Itoh1997SP, Ji2024SP, Beattie2021SPCharges, Choi2015SP, Huang2020surfacestateSPFermipinning}. Our further DFT based Berry phase calculation illustrated the intrinsic polarity and SP, showing that 4H-SiC exhibited SP of 1.8965 $\rm C/m^2$ per unit cell along the direction of [000-1].


Prevalent localized surface states on both C-face and Si-face are identified with spatial projected density of states distribution in Fig. S15(b) and electronic band structure depicted in Fig. S17. It is well-established that in the absence of surface states on a semiconductor, carrier movement occurs at the metal-semiconductor interface upon contact, leading to interfacial band bending \cite{Beattie2021SPCharges, Yang2022reasonofsurfacestateSP, Choi2015SP, Huang2020surfacestateSPFermipinning, Hashimoto2020Fermipinning}. However, the abundant surface states suggest the presence of Fermi level pinning effect at the electrode/4H-SiC interfaces \cite{Hashimoto2020Fermipinning, Kim2018FermiPinning, Kim2019FermiPinning}. Previous studies demonstrated the Fermi level alignment to 4H-SiC surfaces despite variations in electrodes \cite{Ji2024SP, Li2023diffelectrodes}. The energy diagram of the pure 4H-SiC resistors with TiN electrodes illustrates a pinning-type conductor/semiconductor contact on either C-face or Si-face in Fig. 2(b), with a barrier height of 0.14 eV on C-face (0.62 eV on Si-face), which corresponds to the difference measured $W_{F,C}$ (or $W_{F,Si}$) and the work function of interior 4H-SiC. The notable difference in surface potential between the C-face and Si-face eventually induced a potential disparity, consequently causing an inherent bias across the 4H-SiC resistor without any surface electronic state modifications; see in Fig. 2(b).

By introducing ALD ZnO as interlayers into the 4H-SiC resistors, we know that the localized surface states on 4H-SiC surfaces undergo remarkable changes, with their band diagram shown in Fig. 2(c) . DFT calculations predicted the work functions changes with 3 nm ALD ZnO interlayers on C-face from 5.61 eV to 4.29 eV and a slight regulation on Si-face from 4.36 eV to 4.28 eV in Fig. 2(d, e), deviating from the work function of ALD ZnO (4.57 eV) \cite{Beh2017ALD}. These predicted trends of changes after atomic engineering align perfectly with the patterns observed in the experimental results shown in Fig. 1(g). This manipulation of surface electronic states was proven by ALD ZnO from a theoretical perspective. Interfacial coupling shown in band diagrams of Fig. 2(f) induces downward band bending below the Fermi level at the ZnO/C-face interface, while a distinct variation in depth and width of band bending was found for the Si-face in Fig. 2(g). Resorting to the downward band bending below the Fermi level, the freely mobile surface charges are effectively immobilized, allowing constant resistance in the \textit{I-V} characterization. Meanwhile, DFT-based Berry phase calculation predicted ZnO SP of 0.9176 $\rm C/m^2$ per unit cell, along the direction of [000-1], which is lower than that of 4H-SiC (1.8965 $\rm C/m^2$). To completely eliminate the SP-induced bias across the resistor so that the \textit{I-V} curve can cross the origin, the thickness of the ALD-deposited ZnO layer must be meticulously adjusted, as demonstrated in our experimental study.


Moreover, the strong surface states of the additional ALD ZnO interlayers (as evidenced by the upward bending valence in Fig. 2(f, g)) on both faces also suggested the Fermi level pinning effect. At the device level, when using different electrodes or electrode pairs (Fig. S18 and S19 (a-d)) with 3 nm ZnO interlayers on both faces, all configurations show ideal Ohmic properties, as evidenced in Figs. S19 (e) and S19 (f), respectively. This indicates that the ZnO interlayers mitigate work function difference-induced band bending by preventing wavefunction penetration from the conductors to the semiconductors \cite{Hashimoto2020Fermipinning, Kim2018FermiPinning, Kim2019FermiPinning}.
By the Fermi level pinning effect realized with ZnO interlayers, flattened Fermi level arises and internal electric field within the device is absent due to the almost identical work functions, leading to ideal Ohmic \textit{I-V} characteristic with constant resistance regardless of applied voltage in Fig. 2(c).

\begin{figure}
\includegraphics[width=8.6 cm,page=3]{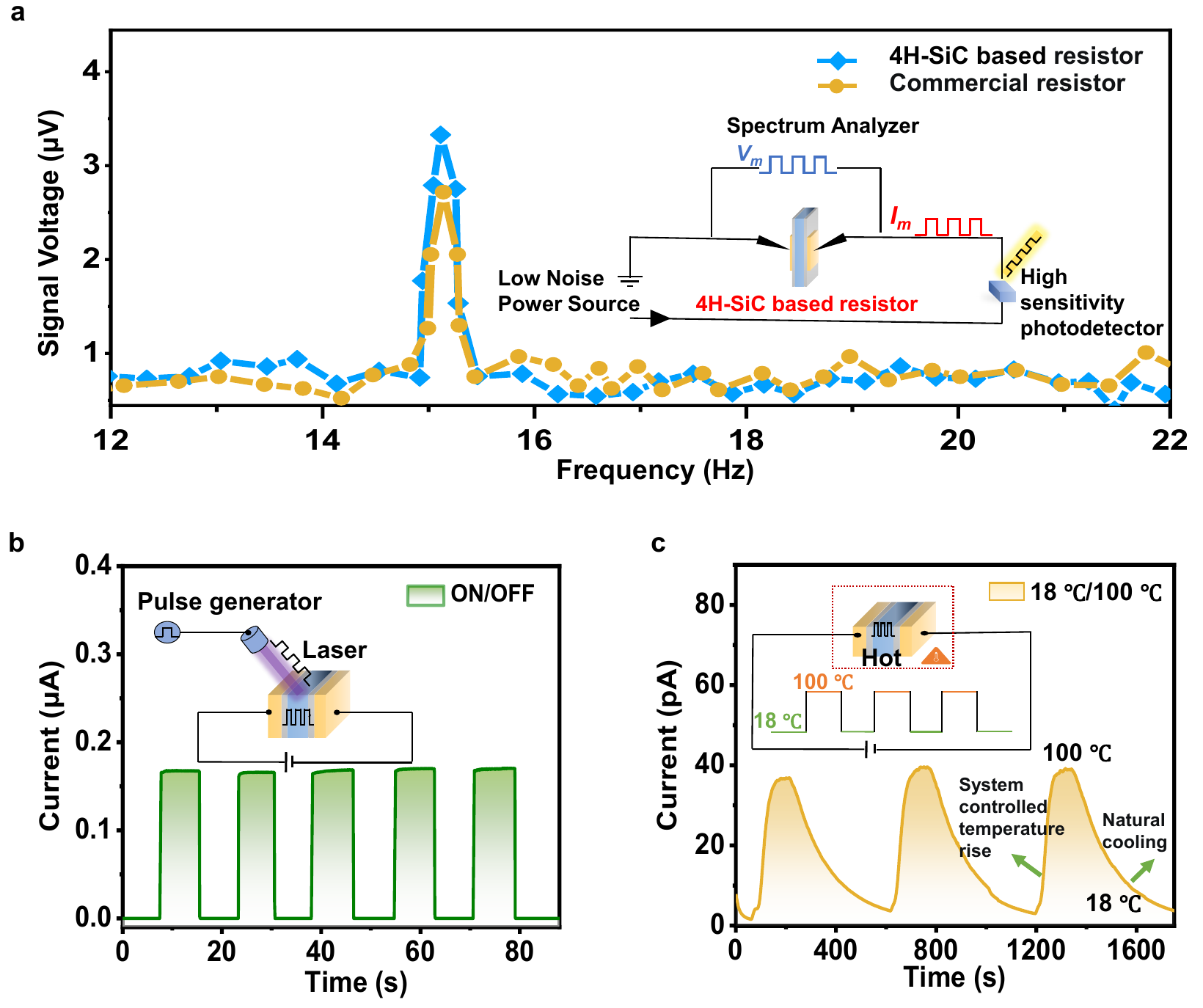}
\caption{\textbf{Applications of 4H-SiC high-resistance resistor.} (a) Weak photocurrent response measurement with 4H-SiC resistor and 10 $\rm T\Omega$ commercial resistor. (inset: schematic diagram of the circuit connection for weak photocurrent response measurement using 4H-SiC resistor.) (b) Photo-sensitive characteristic of 4H-SiC resistor with a periodic light switching. (c) Thermal-sensitive characteristic of 4H-SiC resistor with periodically varying temperature. }
\end{figure}

Finally, we explored the application potential of our ultrahigh-resistance resistors. In general, weak current signals are magnified by resistance, converting minute current to output voltage \cite{Park2020OhmtoVoltage, Wang2021OhmtoVoltage, Scarsella2023OhmtoVoltage}. As illustrated by the schematic model in Fig. 1(a), the ultrahigh resistance with a balanced Fermi level shows potential in the precise measurement industry. 
In Fig. 3(a), in terms of frequency response performance, our 4H-SiC resistors are comparable to the commercial resistors (the highest resistance of 10 $\rm T\Omega$ available in the market, Fig. S3 \cite{sm}) for signal magnification. 
Beyond serving as resistors, these devices can also function as switching elements responding to optical or thermal signals, as shown in Fig. 3(b) and (c). 
Dynamical response to temperature of the devices is shown by a real-time record of resistance measurements in Video S1. 
Moreover, Fig. S20 demonstrates that even minute changes in optical or thermal signals can trigger substantial alterations in resistance values. 
This unique characteristic makes these devices as promising candidates for high performance switching applications in precise sensing technologies.

For the first time, we demonstrate the tuning of peta-ohm-level resistance in semi-insulating 4H-SiC resistors using atomic-scale metal oxide interlayers. We find that freely mobile surface charges generated by dangling bonds and unbalanced surface potentials on pristine C- and Si-faces of 4H-SiC induce nonlinear \textit{I-V} characteristics. In contrast, atomic-layer-deposited ZnO immobilizes these charges and offsets surface polarization, as evidenced by KPFM and DFT simulations. Compared to commercial resistors, our ZnO-modified 4H-SiC devices exhibit exceptional linearity in \textit{ I-V} response, with highly reproducible resistance up to the peta-ohm range over voltages extending to 1000 V. This breakthrough enables high-precision applications requiring extreme resistance stability.

\begin{acknowledgments}
The authors thank Rong Chen and Subrahmanyam Pattamatta for providing insightful guidance in simulation and analysis. This work was supported by National Natural Science Foundation of China (grant U21A20496), The Key Research and Development Program of Shanxi Province (grant 202102150101007), Fund Program for the Scientific Activities of Selected Returned Overseas Professionals in Shanxi Province (grant 20230011), Research Program Supported by Shanxi-Zheda Institute of Advanced Materials and Chemical Engineering (grant 2022SX-TD020), The Central Government Guides Local Funds for Scientific and Technological Development (grant YDZJSX20231A010), HK Jockey Club Charities Trust (grant GSP 181), RGC Strategic Topics Grant (grants STG3/E-704/23-N).
\end{acknowledgments}

\nocite{*}
\bibliography{reference}

\end{document}